\documentclass[12pt]{article}
 \usepackage{epsfig}
 \def\be{\begin{equation}}
 \def\ee{\end{equation}}
 \def\bea{\begin{eqnarray}}
 \def\eea{\end{eqnarray}}
 \usepackage{graphicx}% Include figure files

 \catcode`\@=11
 \def\lsim{\mathrel{\mathpalette\@versim<}}
 \def\gsim{\mathrel{\mathpalette\@versim>}}
 \def\@versim#1#2{\vcenter{\offinterlineskip
 \ialign{$\m@th#1\hfil##\hfil$\crcr#2\crcr\sim\crcr } }}
 \catcode`\@=12

 \catcode`@=12
\begin{document}
 \thispagestyle{empty}
 \begin{flushright}
 UCRHEP-T599\\
 Jul 2019\
 \end{flushright}
 \vspace{0.6in}
 \begin{center}
 {\LARGE \bf Two-Loop $Z_4$ Dirac Neutrino Masses and\\  
Mixing, with Self-Interacting Dark Matter\\}
 \vspace{1.2in}
 {\bf Ernest Ma\\}
 \vspace{0.2in}

{\sl Physics and Astronomy Department,\\ 
University of California, Riverside, California 92521, USA\\}
\vspace{0.1in}
%{\sl Jockey Club Institute for Advanced Study,\\ 
%Hong Kong University of Science and Technology, Hong Kong, China\\} 
\end{center}
 \vspace{1.2in}

\begin{abstract}\
Choosing how gauge $U(1)_\chi$ breaks in the context of 
$SO(10) \to SU(5) \times U(1)_\chi$,\\ $Z_4$ lepton number may be obtained 
which maintains neutrinos as Dirac fermions.  Choosing $\Delta(27)$ as 
the family symmetry of leptons, tree-level Dirac neutrino masses may 
be forbidden.  Choosing a specific set of self-interactimg dark-matter 
particles, Dirac neutrino masses and mixing may then be generated in 
two loops.  This framework allows the realization of cobimaximal 
neutrino mixing, i.e. $\theta_{13} \neq 0$, $\theta_{23} = \pi/4$, 
$\delta_{CP} = \pm \pi/2$, as well as the desirable feature that the 
light scalar mediator of dark-matter interactions decays only to neutrinos, 
thereby not disrupting the cosmic microwave background (CMB).  
\end{abstract}

\newpage
\baselineskip 24pt

\section{Introduction}

The fundamental issue of whether neutrinos are Majorana remains 
open, without any incontrovertible experimental evidence that 
they are so, i.e. no definitive measurement of a nonzero neutrinoless 
double beta decay.  If they are Dirac, for each left-handed $\nu_L$ 
observed in weak interactions, there must be a corresponding right-handed 
$\nu_R$, which has no interactions within the standard model (SM) of quarks 
and leptons.  To justify its existence, the canonical choice is to extend 
the SM gauge symmetry $SU(3)_C \times SU(2)_L \times U(1)_Y$ to the 
left-right symmetry 
$SU(3)_C \times SU(2)_L \times SU(2)_R \times U(1)_{(B-L)/2}$. 
In that case, the $SU(2)_R$ doublet $(\nu,e)_R$ is required, and the 
charged $W_R^\pm$ gauge boson is predicted along with a neutral $Z'$ gauge 
boson.

A more recent choice is to consider $U(1)_\chi$ which comes from 
$SO(10) \to SU(5) \times U(1)_\chi$, with $SU(5)$ breaking to the SM 
at the same grand unified scale.   Assuming that $U(1)_\chi$ survives 
to an intermediate scale, the corresponding $Z_\chi$ gauge boson has 
prescribed couplings to the SM quarks and leptons, which allow current 
experimental data to put a lower bound of about  
4.1 TeV~\cite{atlas-chi-17,cms-chi-18} on its mass. 
In this scenario, $\nu_R$ is a singlet and it exists for the cancellation 
of gauge anomalies involving $U(1)_\chi$.  Using this new framework, 
new insights into dark matter~\cite{m18, m19-1} and Dirac neutrino 
masses~\cite{m19-2,m19-3} have emerged.  

To break $U(1)_\chi$, a singlet scalar is the simplest choice, but it must 
not couple to $\nu_R \nu_R$, or else a Majorana mass for $\nu_R$ would be 
generated.  This simple idea was first discussed~\cite{mpr13} in 2013 
in the general case of singlet fermions charged under a gauge $U(1)_X$.  
If a scalar with three units of $X$ charge is used to break it, 
these fermions with one unit of $X$ charge would not be able ever to 
acquire Majorana masses.  Hence a residual global U(1) symmetry remains.  
This idea is easily applicable to lepton number~\cite{ms15} as well.  

In the SM, the Yukawa couplings linking $\nu_L$ to $\nu_R$ through the 
SM Higgs boson must be very small if neutrinos are Dirac.  
To avoid these tiny tree-level couplings, some additional symmetry is 
often assumed which forbids them.  However, since neutrinos are known to 
have mass, this symmetry cannot be exact.  Indeed, Dirac neutrino masses 
may be generated radiatively as this symmetry is broken softly by 
dimension-three terms.  For a generic discussion, see Ref.~\cite{mp17}, 
which is fashioned after that for Majorana neutrinos~\cite{m98}.  
In some applcations~\cite{gs08,fm12,bmpv16}, the particles in the loop 
belong to the dark sector.  This is called the scotogenic mechanism, 
from the Greek 'scotos' meaning darkness, the original one-loop 
example~\cite{m06} of which was applied to Majorana neutrinos.

Instead of the {\it ad hoc} extra symmetry which forbids the tree-level 
couplings, unconventional assignments of the gauge charges of $\nu_R$ may be 
used~\cite{ms15,yd18,bccps18,cryz18,dkp19} instead.  However, a much more 
attractive idea is to use a non-Abelian discrete family symmetry, which is  
softly broken in the dark sector.  In particular,  
$\Delta(27)$~\cite{m06-1,vkr07,m13-1,abmpv14,m19-4} has been shown to be 
useful in achieving the goal of having scotogenic Dirac neutrino masses 
with a mixing pattern~\cite{bmv03,gl04,mn12} called 
cobimaximal~\cite{m15-1,h15,m16-1,m16-2,fgjl16,gl17}, 
i.e. $\theta_{23} = \pi/4$ and $\delta_{CP} = \pm \pi/2$, which is consistent 
with present neutrino oscillation data~\cite{t2k18} for $\delta_{CP} = -\pi/2$.

\section{Outline of Model}

Following Refs.~\cite{m19-3,m19-4}, the interplay between $U(1)_\chi$ and 
$\Delta(27)$ is used for restricting the interaction terms among the various 
fermions and scalars.  The irreducible representations of $\Delta(27)$ 
and their character table are given in Ref.~\cite{m06-1}.  Note that 
if a set of 3 complex fields transforms as the 3 
representation of $\Delta(27)$, then its conjugate transforms as $3^*$, 
which is distinct from 3.  The basic multiplication rules are
\begin{equation}
3 \times 3 = 3^* + 3^* + 3^*, ~~~ 3 \times 3^* = \sum^9_{i=1} 1_i.
\end{equation}
The particles of this model are shown in Table 1. 
\begin{table}[tbh]
\centering
\begin{tabular}{|c|c|c|c|c|c|c|c|c|}
\hline
particle & $SO(10)$ & $SU(3)_C$ & $SU(2)_L$ & $U(1)_Y$ & 
$U(1)_\chi$ & $\Delta(27)$ & $Z_4^L$ & $R_\chi$ \\
\hline
$(\nu,e)$ & 16 & 1 & 2 & $-1/2$ & 3 & 3 & $i$ & + \\ 
$e^c$ & 16 & 1 & 1 & 1 & $-1$ & $1_1,1_7,1_4$ & $-i$ & + \\ 
$\nu^c$ & 16 & 1 & 1 & 0 & $-5$ & $3^*$ & $-i$ & $+$ \\ 
\hline
$N$ & $126^*$ & 1 & 1 & 0 & 10 & $1$ & $-1$ & $-$ \\ 
$N^c$ & 126 & 1 & 1 & 0 & $-10$ & 1 & $-1$ & $-$ \\ 
$(E_1^0,E_1^-)$ & 10 & 1 & 2 & $-1/2$ & $-2$ & $3^*$ & 1 & $-$ \\ 
$(E_2^+,E_2^0)$ & 10 & 1 & 2 & $1/2$ & 2 & 3 & 1 & $-$ \\ 
$S$ & 45 & 1 & 1 & 0 & 0 & 1 & 1 & $-$ \\
\hline
\hline
$(\phi_1^0,\phi_1^-)$ & 10 & 1 & 2 & $-1/2$ & $-2$ & $3^*$ & 1 & +  \\ 
$(\phi_2^+,\phi_2^0)$ & 10 & 1 & 2 & $1/2$ & 2 & 3 & 1 & + \\ 
\hline
$\sigma$ & 16 & $1$ & 1 & $0$ & $-5$ & 3 & $-i$ & $-$  \\ 
$\zeta_2$ & 126 & 1 & 1 & 0 & $-10$ & 1 & $-1$ & + \\
$\zeta_4$ & 2772 & 1 & 1 & 0 & $-20$ & 1 & 1 & + \\
\hline
\end{tabular}
\caption{Particle content of model.}
\end{table}

In the notation above, all fermion fields are left-handed.  The usual 
right-handed fields are denoted by their charge conjugates.  The SM 
particles transform under $U(1)_\chi$ according to their $SO(10)$ origin, 
as well as the particles of the dark sector $(\sigma, N, N^c, S, E_{1,2})$. 
The input family symmetry is $\Delta(27)$.  The gauge $U(1)_\chi$ is broken 
by $\zeta_4$.  The allowed terms $\zeta_2^2 \zeta_4^*$ and 
$\sigma^2 \zeta_2^*$ 
imply that a residual $Z_4^L$ symmetry~\cite{hr13,h13,cmsv17,csv16,hsv18} 
remains for lepton number as shown 
in Table 1.  The dark symmetry is simply $R_\chi = (-1)^{Q_\chi + 2j}$ as 
pointed out recently~\cite{m18}.  Note that the dark scalar $\sigma$ is 
also a lepton~\cite{m15} because it has the same $Z_4^L$ charge as $\nu^c$.   
The complete Lagrangian is invariant under gauge $U(1)_\chi$ in all 
its terms, as well as $\Delta(27)$ in all the dimension-four terms. 
Whereas the breaking of gauge $U(1)_\chi$ must only be spontaneous,  
through the vacuum expectation values of $\zeta_4$ and $\Phi_{1,2}$, 
the breaking of $\Delta(27)$ is both spontaneous, through the vacuum 
expectation values of $\Phi_{1,2}$, and explicit, through the soft 
dimension-three terms $\sigma_j \sigma_k \zeta_2^*$, as shown below. 

From Table 1, the Yukawa term $e e^c \phi_1^0$ is allowed, but $\Delta(27)$ 
forbids $\nu \nu^c \phi_2^0$, hence neutrinos do not have tree-level Dirac 
masses.  Moreover, the usual dimension-five operator for Majorana 
neutrino mass, i.e. $\nu \nu \phi_2^0 \phi_2^0$, is forbidden as well 
as the usual singlet Majorana mass term $\nu^c \nu^c$.  Without 
$U(1)_\chi$, $\nu^c \nu^c$ would be a soft term breaking $\Delta(27)$ 
and would then have been allowed by itself.  To obtain Dirac neutrino 
masses, the fermion doublets $E_{1,2}$ and singlets $S,N,N^c$ with even 
$Q_\chi$ as well as the scalar singlet $\sigma$ with odd $Q_\chi$ are added.  
They belong to the dark sector because SM fermions have odd $Q_\chi$ and 
the SM Higgs doublet has even $Q_\chi$, as explained in Ref.~\cite{m18}. 
With the above particle content, Dirac neutrino masses cannot be generated  
in one loop, but are possible in two loops with the soft breaking trilinear 
couplings of $\sigma_j \sigma_k \zeta_2^*$, as shown in Fig.~1.
\begin{figure}[htb]
\vspace*{-5cm}
\hspace*{-3cm}
\includegraphics[scale=1.0]{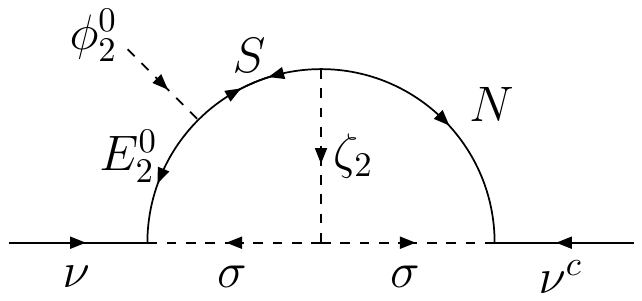}
\vspace*{-21.5cm}
\caption{Two-loop diagram for scotogenic $U(1)_\chi$ Dirac neutrino 
masses and mixing.}
\end{figure}

In the above, only $\phi^0_2$ is shown, but it can be replaced by 
$\bar{\phi}^0_1$.  Since $\phi_2^0$ is a 3 under $\Delta(27)$, its 
components are denoted as $\phi^0_{2,i}$ with $i=1,2,3$. 
The dark scalars and fermions have allowed interactions with $\nu,\nu^c$ 
under $U(1)_\chi$.  The dimension-four terms, i.e. $\nu \sigma E^0_2$, 
$\nu^c \sigma N$, $E_2^0 \bar{\phi}_2^0 S$, $S N \zeta_2$, respect 
both $U(1)_\chi$ and $\Delta(27)$.  The dimension-three scalar trilinear 
couplings $\sigma_j \sigma_k \zeta_2^*$ respect $U(1)_\chi$ but not 
$\Delta(27)$.

Consider now the spontaneous breaking of $U(1)_\chi$.  First, because 
$\nu^c \sim 3^*$ under $\Delta(27)$ and has $Q_\chi = -5$, it is protected 
from acquiring a tree-level Majorana mass.  Choosing $\zeta_4$ (instead 
of $\zeta_2$) to have a nonzero vacuum expectation value then makes 
the residual lepton symmetry $Z_4^L$, through the allowed couplings 
$\zeta_2^2 \zeta_4^*$ and $\sigma^2 \zeta_2^*$, with their connections to 
leptons from $\nu^c \sigma N$ and $\nu \sigma E^0_2$.

\section{Neutrino Mixing}

Using the decomposition $3 \times 3^*$ and 
$\langle \phi^0_{2,i} \rangle = v_i$, 
with $1_1,1_7,1_4$ as defined in Ref.~\cite{m06-1}, instead of the usual 
$1_1,1_2,1_3$ of the original $A_4$ model~\cite{mr01} of neutrino mixing,  
the charged-lepton mass matrix is given by
\begin{equation}
{\cal M}_l = \pmatrix{f_e v_1^* & f_\mu v_3^* & f_\tau v_2^* \cr 
f_e v_2^* & f_\mu v_1^* & f_\tau v_3^* \cr 
f_e v_3^* & f_\mu v_2^* & f_\tau  v_1^*} = 
\pmatrix{m_e & 0 & 0 \cr 0 & m_\mu & 0 \cr 0 & 0 & m_\tau},
\end{equation}
where $v_2=v_3=0$ has been assumed for the spontaneous breaking of 
$\phi^0_2$ (or $\bar{\phi}^0_1$).  This ${\cal M}_l$ is diagonal and 
different from that of Ref.~\cite{mr01}.  
It allows also three independent masses for the charged leptons, and 
the emergence of lepton flavor triality~\cite{m10,cdmw11} in the 
Yukawa interactions of the three charged leptons with the three 
Higgs doublets.

The $\nu \sigma E_2^0$ couplings obey $\Delta(27)$ 
according to 
\begin{equation}
(3 \times 3) \times 3 = (3^* + 3^* + 3^*) \times 3 = 1 + 1 + 1.
\end{equation} 
The three $\Delta(27)$ invariants are
\begin{equation}
111+222+333, ~~ 123+231+312, ~~ 132+321+213.
\end{equation}
However, since only $\phi^0_{2,1}$ has a nonzero vacuum expectation value, 
only $E^0_{2,1}$ matters in the above.  Hence only the 111, 231, and 321 
couplings contribute to the radiative neutrino mass matrix of Fig.~1.  
Because the charged-lepton mass matrix is diagonal, all three couplings 
may be chosen real by 
absorbing their phases.  One magnitude may also be arbitrarily chosen. 
The coupling matrix linking $\nu_i$ to $\sigma_j$ is then
\begin{equation}
\pmatrix{a & 0 & 0 \cr 0 & 0 & c \cr 0 & s & 0},
\end{equation}
where $c^2+s^2=1$. 
The soft breaking of $\Delta(27)$ occurs at the $\sigma_j \sigma_k \zeta_2^*$ 
trilinear vertex.  Choosing the residual $1-1$ and $2-3$ exchange symmetry
with complex conjugation~\cite{gl04}, this $3 \times 3$ coupling matrix 
is of the form
\begin{equation}
\pmatrix{b & e & e^* \cr e & d & f \cr e^* & f & d^*},
\end{equation}
where $b,f$ are real.  The resulting Dirac neutrino mass matrix is 
proportional to their product
\begin{equation}
{\cal M} = \pmatrix{a & 0 & 0 \cr 0 & 0 & c \cr 0 & s & 0} 
\pmatrix{b & e & e^* \cr e & d & f \cr e^* & f & d^*} = 
\pmatrix{ab & ae & ae^* \cr ce^* & cf & cd^* \cr se & sd & sf}.
\end{equation}
This is diagonalized on the left by the unitary neutrino mixing matrix 
$U_{l \nu}$, which may be obtained by considering the Hermitian matrix 
\begin{equation}
{\cal M}{\cal M}^\dagger = \pmatrix{a^2(b^2+2|e|^2) & ac(be+fe+de^*) & 
as(be^*+fe^*+d^*e) \cr ac(be^*+fe^*+d^*e) & c^2(|e|^2+f^2+|d|^2) & 
sc({e^*}^2+2fd^*) \cr as(be+fe+de^*) & sc(e^2+2fd) & s^2(|e|^2+|d|^2+f^2)}.
\end{equation}
Rewriting
\begin{eqnarray}
{\cal M}{\cal M}^\dagger &=& \pmatrix{A & \sqrt{2} c |D|e^{i \theta_D} & 
\sqrt{2} s |D|e^{-i \theta_D} \cr \sqrt{2} c |D| e^{-i \theta_D} & 2c^2 B & 
-2sc |E|e^{2i \theta_E} \cr \sqrt{2} s |D| e^{i \theta_D} & -2sc |E|e^{-2i \theta_E} 
& 2s^2 B} \\ &=& \pmatrix{1 & 0 & 0 \cr 0 & e^{i\theta_E} & 0 
\cr 0 & 0 & e^{-i\theta_E}} \pmatrix{A & \sqrt{2} c |D|e^{i\theta} & 
\sqrt{2} s |D|e^{-i\theta} \cr \sqrt{2} c |D|e^{-i\theta} & 2c^2 B & -2sc |E| 
\cr \sqrt{2} s |D|e^{i\theta} & -2sc |E| & 2s^2 B} \pmatrix{1 & 0 & 0 
\cr 0 & e^{-i\theta_E} & 0 \cr 0 & 0 & e^{i\theta_E}}, \nonumber
\end{eqnarray}
and removing the diagonal phases on both sides, where 
$\theta = \theta_D + \theta_E$, the mass-squared matrix becomes
\begin{equation}
{\cal M}{\cal M}^\dagger = \pmatrix{A & \sqrt{2} c D & \sqrt{2} s D^* \cr 
\sqrt{2} c D^* & 2c^2 B & -2sc E \cr \sqrt{2} s D & -2sc E & 2s^2 B},
\end{equation}
where
\begin{eqnarray}
&& A = a^2(b^2 + 2|e|^2), ~~ B = (1/2)(|d|^2 + |e|^2 + f^2), \\ 
&& D = (a/\sqrt{2})|be+fe+de^*|e^{i\theta}, ~~ E = (1/2)|e^2+2fd|.
\end{eqnarray}
If $c=s=1/\sqrt{2}$, then ${\cal M}{\cal M}^\dagger$ is diagonalized by 
a cobimaximal $U_{l\nu}$, as shown below.

Multiplying Eq.~(10) with $c=s=1/\sqrt{2}$ on the left by 
\begin{equation}
U_{TBM}^\dagger = \pmatrix{\sqrt{2/3} & -1/\sqrt{6} & -1/\sqrt{6} \cr 
1/\sqrt{3} & 1/\sqrt{3} & 1/\sqrt{3} \cr 0 & i/\sqrt{2} & -i/\sqrt{2}} 
\end{equation}
and on the right by $U_{TBM}$, a real matrix is obtained, with 
\begin{eqnarray}
&& M^2_{11} = {1 \over 3}(2A-4D_R+B-E), ~~ M^2_{22} = {1 \over 3} 
(A+4D_R+2B-2E),  \\ 
&& M^2_{12} = M^2_{21} = {\sqrt{2} \over 3}(A+D_R-B+E), ~~ \\ 
&& M^2_{33} = B+E, ~~ M^2_{13} = M^2_{31} = 
{2 D_I \over \sqrt{3}}, ~~ M^2_{23} = M^2_{32} = {\sqrt{2} D_I \over \sqrt{3}}.
\end{eqnarray}
Since a real matrix is diagonalized by an orthogonal matrix ${\cal O}$, 
the product
\begin{equation}
U_{l \nu} = U_{TBM} {\cal O} 
\end{equation}
is easily shown to have the property of $|U_{\mu i}| = |U_{\tau i}|$ for 
$i=1,2,3$, which is the necessary and sufficient condition for 
cobimaximal mixing, i.e. $\theta_{13} \neq 0$, $\theta_{23} = \pi/4$, 
and $\delta_{CP} = \pm \pi/2$.

Using the fact that $U_{TBM}$ is a good approximation of the experimental 
data, the orthogonal matrix may be written as
\begin{equation}
{\cal O} = \pmatrix{1 & s_1 & s_2 \cr -s_1 & 1 & s_3 \cr -s_2 & -s_3 & 1},
\end{equation}
where
\begin{eqnarray}
s_1 &=& {M^2_{12} \over M^2_{22}-M^2_{11}} = {-\sqrt{2}F \over 3D_R-F}, \\ 
s_2 &=& {M^2_{13} \over M^2_{33}-M^2_{11}} = {2D_I \over \sqrt{3}(B+E-A+D_R+F)}, \\ 
s_3 &=& {M^2_{23} \over M^2_{33}-M^2_{22}} = {\sqrt{2}D_I \over 
\sqrt{3}(B+E-A-2D_R+2F)} \simeq {s_2 \over \sqrt{2}},
\end{eqnarray}
with $F=(A+D_R-B+E)/3$.  Since $M^2_{22}-M^2_{11} \simeq 7.37 \times 10^{-5}$ 
eV$^2$ and $M^2_{33}-M^2_{11} \simeq 2.56 \times 10^{-3}$ eV$^2$, the above 
implies $|F| << D_R << |B+E-A|$.  Now
\begin{equation}
U_{l \nu} = \pmatrix{\sqrt{2/3}(1-s_1/\sqrt{2}) & \sqrt{1/3}(1+\sqrt{2}s_1) & 
\sqrt{1/3}(\sqrt{2}s_2+s_3) \cr -\sqrt{1/6}(1+\sqrt{2}s_1)+is_2/\sqrt{2} & 
\sqrt{1/3}(1-s_1/\sqrt{2}) + is_3/\sqrt{2} & -i/\sqrt{2} \cr 
-\sqrt{1/6}(1+\sqrt{2}s_1)-is_2/\sqrt{2} & 
\sqrt{1/3}(1-s_1/\sqrt{2}) - is_3/\sqrt{2} & i/\sqrt{2}}.
\end{equation}
Note first that $U_{\tau i} = U_{\mu i}^*$.  Multiplying the third row by $-1$ 
and the third column by $i$, the PDG convention of $U_{l \nu}$ is obtained, 
with $\delta_{CP} = - \pi/2$ for
\begin{equation}
s_{13} =  {\sqrt{2} s_2 \over \sqrt{3}} + {s_3 \over \sqrt{3}} = 
{\sqrt{2}D_I \over \sqrt{3}(B+E-A)} > 0.
\end{equation}
At the same time, $\theta_{23} = \pi/4$, and
\begin{equation}
\tan \theta_{12} = {1+\sqrt{2}s_1 \over \sqrt{2}(1-s_1/\sqrt{2})}.
\end{equation}
Using $\sin^2 \theta_{12} = 0.297$, the above implies $s_1 = -0.039$. 
Note that the deviations from $\tan^2 \theta_{12} = 1/2$ due to $s_2$ 
and $s_3$ are quadratic.  For $s_{13}^2=0.0215$, their contributions 
shift $s_1$ to $-0.041$.

To see how deviations from cobimaximal mixing occur, let 
$\sqrt{2}s - 1 = \epsilon$ and $\sqrt{2}c - 1 = -\epsilon$, then
\begin{equation}
\Delta ({\cal M}{\cal M}^\dagger) = \pmatrix{0 & -\epsilon D & \epsilon D^* 
\cr -\epsilon D^* & -2\epsilon B & 0 \cr \epsilon D & 0 & 2\epsilon B}.
\end{equation}
Multiplying on the left by $U^\dagger_{TBM}$ and on the right by $U_{TBM}$, 
this becomes
\begin{equation}
i \epsilon \pmatrix{0 & -\sqrt{2} D_I & -(2/\sqrt{3})(B-D_R) \cr 
\sqrt{2}D_I & 0 & \sqrt{2/3} (2B+D_R) \cr (2/\sqrt{3})(B-D_R) & 
-\sqrt{2/3}(2B+D_R) & 0 }.
\end{equation}
The additional mixing contributions analogous to $s_{1,2,3}$ are thus
\begin{eqnarray}
i \epsilon_1 &=& {-i \epsilon \sqrt{2} D_I \over 3D_R-F}, \\ 
i \epsilon_2 &=& {-2 i \epsilon (B-D_R) \over 3(B+E-A+D_R+F)}, \\ 
i \epsilon_3 &=& {\sqrt{2} i \epsilon (2B+D_R) \over 3(B+E-A-2D_R+2F)} \simeq 
-\sqrt{2} i \epsilon_2.
\end{eqnarray}
Numerically, $\epsilon_1$ is enhanced by 
$\sqrt{3} s_{13} \Delta m_{31}^2/\Delta m_{21}^2 \simeq 8.82$, but not 
$\epsilon_{2,3}$.  Hence the rotation matrix of Eq.~(18) due to $s_{1,2,3}$ 
is replaced with 
\begin{equation}
U_\epsilon = \pmatrix{1-\epsilon_1^2/2 & s_1+i\epsilon_1 & s_2+i\epsilon_2 
\cr -s_1+i\epsilon_1 & 1- \epsilon_1^2/2 & s_3+i\epsilon_3 \cr 
-s_2+i\epsilon_2 & -s_3+i\epsilon_3 & 1},   
\end{equation}
and $U_{l \nu} = U_{TBM} U_\epsilon$ instead of Eq.~(17).  The various 
entries of $U_{l \nu}$ are thus
\begin{eqnarray}
&& U_{e3} = s_{13}, ~~ U_{\mu 3} = {-i \over \sqrt{2}} \left(1- 
\sqrt{3 \over 2} \epsilon_3 \right), \\ 
&& U_{e2} = {1 \over \sqrt{3}} \left(1-{\epsilon_1^2 \over 2} 
+ \sqrt{2}s_1 \right) + i \sqrt{2 \over 3} \epsilon_1, \\ 
&& U_{\mu 2} = {1 \over \sqrt{3}} \left( 1 - {\epsilon_1^2 \over 2} - 
{s_1 \over \sqrt{2}} + \sqrt{3 \over 2} \epsilon_3 \right)  
- {i \epsilon_1 \over \sqrt{6}}.
\end{eqnarray}
To obtain $\delta_{CP}$, the identity
\begin{equation}
J_{CP} = Im(U_{\mu 3} U^*_{e3} U_{e2} U^*_{\mu 2}) = 
c_{13} s_{12} c_{12} s_{23} c_{23} s_{13} c_{13} \sin \delta_{CP}
\end{equation}
is used, where 
\begin{eqnarray}
&& s_{23} = {1 \over \sqrt{2}}\left( 1 - \sqrt{3 \over 2} \epsilon_3 \right), 
~~ c_{23} = {1 \over \sqrt{2}}\left( 1 + \sqrt{3 \over 2} \epsilon_3 \right), 
\\ 
&& s_{12} = {1 \over \sqrt{3}} \left( 1 + \sqrt{2}s_1 + {\epsilon_1^2 \over 2} 
\right), ~~  
c_{12} = \sqrt{2 \over 3} \left( 1 + {s_1 \over \sqrt{2}} - 
{\epsilon_1^2 \over 4} \right).
\end{eqnarray}
This implies
\begin{equation}
\sin \delta_{CP} = - \left( 1 - {7\epsilon_1^2 \over 4} \right).
\end{equation}
Using $B=A+E+D_R-3F$, the deviations of $s_{23}^2$ and $\sin \delta_{CP}$ 
from 1/2 and $-1$ are given by
\begin{equation}
1-2s_{23}^2 = {4 \over \sqrt{3}} \left( {A+E \over 2E} \right) \epsilon, 
~~~  1+\sin \delta_{CP} = {7 \over 4} (8.82)^2 \epsilon^2.
\end{equation}
As an example, for $\epsilon=0.02$ and $A=E$, $s^2_{23}=0.48$ and 
$\sin \delta_{CP} = -0.95$.

\section{Dark Sector}

In this two-loop model, the scalar $\sigma$ is a pure singlet.  This means 
that it interacts with quarks not through the $Z$ boson, but rather the 
$Z_\chi$ gauge boson.  The lightest of $\sigma_{1,2,3}$ is dark matter.  Its 
annihilation to the light scalar mediator $\zeta_2$ 
is a well-known mechanism for generating the correct dark-matter relic 
abundance of the Universe.

At the mass of 150 GeV, the constraint on the elastic scattering cross 
section of $\sigma_1$ off nuclei per nucleon is about 
$1.5 \times 10^{-46}$~cm$^2$ from 
the latest XENON result~\cite{xenon18}.  This puts a lower limit on the 
mass of $Z_\chi$, i.e. 
\begin{equation}
\sigma_0 = {\mu_\sigma^2 \over 64\pi} {[Z f_P +(A-Z) f_N]^2 \over A^2} 
< 1.5 \times 10^{-10}~{\rm pb},
\end{equation}
where $\mu_\sigma$ is the reduced mass of $\sigma_1$, and 
\begin{equation}
f_P = g^2_{Z_\chi} f_\sigma (2u_V + d_V)/M^2_{Z_\chi}, ~~~ 
f_N = g^2_{Z_\chi} f_\sigma (u_V + 2d_V)/M^2_{Z_\chi}, ~~~ 
\end{equation}
and $Z=54$, $A=131$ for xenon.  In $U(1)_\chi$, the vector couplings are 
\begin{equation}
f_\sigma = -\sqrt{5 \over 8}, ~~~ u_V =0, ~~~ d_V = {-1 \over \sqrt{10}}.
\end{equation}
Using $\alpha_\chi = g^2_{Z_\chi}/4\pi = 0.0154$ from Ref.~\cite{m18}, 
the bound $M_{Z_\chi} > 24$ TeV is obtained.  

Because of the $\zeta_2^* \sigma_1 \sigma_1$ interaction, $\sigma_1$ is 
a self-interacting dark-matter (SIDM) candidate~\cite{kkpy17} which has 
been postulated to explain the flatness of the core density profile of 
dwarf galaxies~\cite{detal09} and other related astrophysical phenomena.  
The light scalar mediator $\zeta_2$ transforms as $-1$ under $Z_4^L$ 
lepton symmetry and decays only to $\nu^c \nu^c$ in one loop as shown in 
Fig.~2, where the allowed Majorana mass term $NN \langle \zeta_4 \rangle$ 
has been used. 
\begin{figure}[htb]
\vspace*{-5cm}
\hspace*{-3cm}
\includegraphics[scale=1.0]{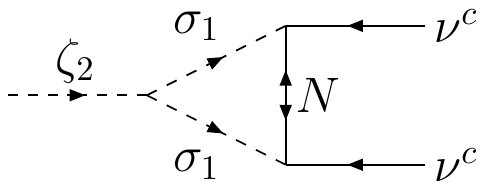}
\vspace*{-21.5cm}
\caption{One-loop diagram for $\zeta_2$ decay to two neutrinos.}
\end{figure}
It does not disrupt~\cite{gibm09} 
the cosmic microwave background (CMB)~\cite{planck16}, thus eluding the 
stringent constraint~\cite{bksw17} due to the enhanced Sommerfeld production 
of $\zeta_2$ at late times if it decays to electrons and photons, as in 
most proposed models.  This problem may also be solved if the light mediator 
is stable~\cite{m17,m18-2,dsw18} or if it decays into $\nu \nu$ through a 
pseudo-Majoron in the singlet-triplet model of neutrino mass~\cite{mm17}. 
A much more natural solution is for it to decay into $\nu^c\nu^c$ as 
first pointed out in the prototype model of Ref.~\cite{m18-1} and 
elaborated in Refs.~\cite{m18,m19-2,m19-3}.   Here it is shown how it may 
arise in the scotogenic Dirac neutrino context using $U(1)_\chi$ as well as 
$\Delta(27)$.  The generic connection of lepton parity  
to simple models of dark matter was first pointed out in Ref.~\cite{m15}. 
Typical mass ranges for $\sigma_1$ and $\zeta_2$ are
\begin{equation}
100 < m_\sigma < 200~{\rm GeV}, ~~~ 10 < m_{\zeta'} < 100~{\rm MeV}, 
\end{equation}
as shown in Ref.~\cite{m18-1}, where details of relic abundance and the 
required elastic cross section for SIDM are explicitly given.

\section{Concluding Remarks}

A recent insight concerning lepton number symmetry is that it could be 
$Z_N$ with $N \neq 2$.  This paper shows explicitly a model with $Z_4^L$ 
lepton symmetry in the context of 
$SU(3)_C \times SU(2)_L \times U(1)_Y \times U(1)_\chi$, where $U(1)_\chi$ 
comes from $SO(10) \to SU(5) \times U(1)_\chi$, and the non-Abelian discrete 
symmetry $\Delta(27)$ as its family symmetry.  With the particle content 
of Table~1, where $U(1)_\chi$ is spontaneously broken by $\zeta_4$ and 
$\Delta(27)$ explicitly broken by the soft trilinear 
$\sigma_j \sigma_k \zeta_2^*$ scalar vertex, Dirac neutrino masses are 
radiatively generated in two loops through the dark sector, which 
consists of particles odd under $R_\chi = (-1)^{Q_\chi + 2j}$.  A pattern of 
neutrino mixing is obtained which fits the cobimaximal hypothesis, i.e. 
$\theta_{13} \neq 0$, $\theta_{23}=\pi/4$, $\delta_{CP} = \pm \pi/2$, with 
possible deviations shown in Eq.~(38).  The lightest of the scalar singlets 
$\sigma_{1,2,3}$ is self-interacting dark matter, with $\zeta_2$ as its light 
scalar mediator which decays only to two neutrinos.

\section{Acknowledgement}

This work was supported in part by the U.~S.~Department of Energy Grant 
No. DE-SC0008541.  

\baselineskip 18pt

\bibliographystyle{unsrt}

\end{document}